\documentclass[manuscript]{acmart}
\AtBeginDocument{%
  }

\setcopyright{acmlicensed}
\copyrightyear{2018}
\acmYear{2018}
\acmDOI{XXXXXXX.XXXXXXX}
\acmConference[OzCHI 2025]{OzCHI 2025}{29 November--3 December, 2025}{Sydney, Australia}
\acmISBN{978-1-4503-XXXX-X/2018/06}

\begin{document}

\title{Designing Drone Interfaces to Assist Pedestrians Crossing Non-Signalised Roads}

\author{Guixiang Zhang}
\email{singkit7b7@gmail.com}
\orcid{0009-0007-2483-1115}
\affiliation{%
  \institution{Design Lab, The University of Sydney}
  \city{Sydney}
  \state{NSW}
  \country{Australia}
}

\author{Yiyuan Wang}
\email{yiyuan.wang@sydney.edu.au}
\orcid{0000-0003-2610-1283}
\affiliation{%
  \institution{Design Lab, The University of Sydney}
  \city{Sydney}
  \state{NSW}
  \country{Australia}
}

\author{Marius Hoggenmüller}
\email{marius.hoggenmueller@sydney.edu.au}
\orcid{0000-0002-8893-5729}
\affiliation{%
  \institution{Design Lab, The University of Sydney}
  \city{Sydney}
  \state{NSW}
  \country{Australia}
}

\renewcommand{\shortauthors}{Zhang et al.}

\begin{abstract}
Recent research highlights the potential of drones to enhance pedestrian experiences, such as aiding navigation and supporting street-level activities. This paper explores the design of drone interfaces to assist pedestrians crossing dangerous roads without designated crosswalks or traffic lights, leveraging drones' ability to monitor and analyse real-time traffic data. Inspired by existing traffic signal systems, the interface communicates safety information through permissive alerts, prohibitive warnings, directional warnings, and collision emergency warnings. These safety cues were integrated into drone interfaces using in-situ projections and drone-equipped screens through an iterative design process. A mixed-methods, within-subjects VR evaluation (n=18) revealed that drone-assisted systems significantly improved pedestrian safety experiences and reduced mental workload compared to a baseline without any crossing aid, with projections outperforming screens. The findings suggest the potential for drone interfaces to be integrated into connected traffic systems. We also offer design recommendations for developing drone interfaces that support safe pedestrian crossings.
\end{abstract}


\maketitle

\section{Introduction}
Unmanned aerial vehicles (UAVs) known as drones are becoming increasingly prevalent in public urban spaces, encompassing various applications for both recreational and professional purposes such as photography, delivery services, surveillance, and search-and-rescue operations~\cite{herdel2021public}. Research studies have also explored speculative design concepts aimed at expanding the use of drones to enhance urban living experiences. For example, drones have been envisioned as `companions' that assist pedestrians in safely returning home at night, reducing anxiety and improving perceived safety~\cite{kim2016getting}. Other concepts include facilitating running groups~\cite{baldursson2021drorun}, in-situ navigation~\cite{knierim2018quadcopter}, and enabling street games~\cite{kljun2015streetgamez}, among others. In light of technological advancements, including capabilities such as traffic and crowds monitoring, real-time or predictive traffic data collection and analysis, optimisation of utility networks, battery recharging, and power consumption~\cite{abbas2023survey,muhlhauser2020street,hoggenmueller2019enhancing}, it is anticipated that in the future, drones will harness these innovations and become essential mobile devices, communicating with other smart objects, such as autonomous robots and vehicles, and improving transportation systems within the framework of smart cities and city-wide internet of things (IoT) infrastructure~\cite{herdel2021public,hollander2020save,abbas2023survey,nguyen2019designing}.

The integration of drones into smart cities offers new opportunities for addressing traffic-related challenges~\cite{abbas2023survey,hollander2019investigating}. One pressing concern is pedestrian road safety~\cite{tiwari2020progress,trafficSafety2023}, with over 270,000 fatalities worldwide each year—22\% of all road traffic deaths. Pedestrian safety is particularly critical on uncontrolled roads due to the absence of pedestrian-dedicated facilities such as traffic lights or crosswalks. While literature on human-drone interaction (HDI) has proposed design concepts across a wide range of urban application areas~\cite{kim2016getting,baldursson2021drorun,knierim2018quadcopter,kljun2015streetgamez}, there is a gap in exploring their use for pedestrians in interacting with traffic. At the same time, research on pedestrian-vehicle interaction has increasingly focused on external human-machine interfaces (eHMIs) in anticipation of highly automated vehicles (AVs)~\cite{chang2018video,mahadevan2018communicating,habibovic2018communicating}. These eHMIs -- typically attached to the vehicle~\cite{dey2020taming} -- convey safety-related information and assist pedestrians during street crossings, using communication modalities such as projections~\cite{nguyen2019designing}, gaze~\cite{chang2017eyes}, textual displays~\cite{eisma2021external}, coloured lights~\cite{dey2020color}, and auditory prompts~\cite{pelikan2023designing}. Drone interfaces not only employ similar communication technologies as eHMIs~\cite{baldursson2021drorun,kljun2015streetgamez,scheible2016situ,schneegass2014midair}, but with the arrival of vehicle-to-everything (V2X) technologies and IoT, they also hold the potential to become an integral part of this intelligent, interconnected network~\cite{kavas2022v2x}. Utilising their inherent advantages, including bird-eye view monitoring and flexible mobility~\cite{knierim2018quadcopter}, drones present a promising opportunity to act as dynamic mediators between pedestrians and vehicles~\cite{hoggenmueller2019enhancing,kavas2022v2x}, mitigating safety risks and enhancing the overall urban technology experience.

Leveraging the potential of drones to monitor and analyse real-time traffic data, this paper proposes drone interfaces for assisting pedestrians in interacting with traffic. Specifically, we aim to address the challenge of crossing safety-critical roads and streets in the absence of crosswalks or traffic signals. Following an iterative design process, we created two interface design concepts~(drone with in-situ projections and drone-equipped screens) to provide crossing-related safety cues to pedestrians. To evaluate these interfaces with users in a realistic yet safe manner, we developed a virtual reality (VR) simulation using 360-degree camera recordings of a main road in a busy urban neighbourhood, overlaid with 3D models of the proposed drone interfaces. In a VR evaluation with eighteen participants, the drone-assisted systems significantly improved perceived safety and reduced mental workload during road crossings, demonstrating potential to function as part of connected traffic systems.

\section{Related Work}

\subsection{Pedestrian Road-Crossing Decision and eHMIs}

The pedestrians' decision-making for street crossing is predominantly dependent on street-crossing facilities such as traffic lights and crosswalks, as well as visual and auditory traffic cues~\cite{hollander2020save}. These cues encompass various indicators, such as the status of vehicles (e.g., acceleration, constant speed, deceleration, or braking), the distance gap between vehicles~\cite{lobjois2007age,habibovic2018communicating}, traffic density~\cite{rasouli2019autonomous}, time to collision~\cite{habibovic2018communicating}, and other pertinent factors. Overall, the acquisition of traffic safety-related information influences a pedestrian's inclination to cross the street~\cite{li2018cross}.

With the increasing automation level in vehicles, eHMIs provides new interaction paradigms between pedestrians and AVs~\cite{dey2020taming,rasouli2019autonomous}. Studies found that eHMIs show promise in improving crossing decision-making, safety experiences, and trust in pedestrian-AV interaction~\cite{mahadevan2018communicating,m2021calibrating,habibovic2018communicating}. Interface modalities, such as vehicle-mounted displays~\cite{bazilinskyy2019survey}, projections~\cite{nguyen2019designing}, flashing lights~\cite{dey2020color}, robotic eyes~\cite{chang2017eyes}, and sounds~\cite{pelikan2023designing}, have yielded advantages in their own rights for moderating pedestrian crossing behaviours, depending on situational~\cite{lee2023safe}, environmental~\cite{fratini2023ranking}, or cultural contexts~\cite{lanzer2020designing}, as well as the desired interaction. By serving as a communication medium between AVs and pedestrians, eHMIs enable pedestrians to infer AVs' intent and awareness~\cite{mahadevan2018communicating,chang2017eyes}, leading to safer, more efficient, and pleasant road-crossing experiences~\cite{m2021calibrating}.

\subsection{Public Drone Interaction and Pedestrian Safety}
Although empirical evidence on the use of drones in pedestrian crossing scenarios is currently lacking, with no existing prototypes or actual findings available, several papers have proposed that drones could serve as valuable aids in enhancing pedestrian safety. Examples include the projection of traffic information for street crossings~\cite{hoggenmueller2019enhancing}, connected smartwatches with text for late-night walk~\cite{kim2016getting}, mid-air displays such as screens to convey useful safety information~\cite{schneegass2014midair}, contextual material like police uniforms to denote safety purposes~\cite{herdel2021public}, drone movements to guide pedestrians to safer paths~\cite{colley2017investigating}, and projected navigational cues in response to pedestrian gestures to function as a tour guide~\cite{Cauchard2019}.

Pedestrian safety assistance from AVs often relies on eHMIs. Recent advancements include infrastructure-based eHMI solutions, such as utilising roadside poles~\cite{chauhan2023fostering}, smart curbs~\cite{hollander2022take}, and virtual traffic lights~\cite{martins2019towards}, furthering the V2X vision within the IoT framework. Drones offer significant potential for enhancing public safety due to their mobility, swarm capabilities, and applications in traffic and crowd management~\cite{hildmann2019using,abbas2023survey,bastani2021skyquery,kim2017stinuum}. Therefore, we propose drone-assisted systems as part of the future V2X infrastructure, focused on delivering alerts and warnings for pedestrians in areas without crossing facilities. By acting as dynamic mediators on non-signalised roads, drones provide safety-related cues tailored to pedestrian crossing needs.

\section{Design Process}
At the outset of our design process, the focus was on creating a drone-based interface concept that presents crucial safety information to pedestrians when crossing busy streets without any designated crossing facilities (e.g., crosswalks, traffic lights). To do so, we followed an iterative design approach, incorporating feedback from domain experts to refine the design concept progressively. Two drone-assisted systems were developed throughout the design process: one using spatial visualisations through in-situ projection and the other using text displayed on a drone-equipped screen.

\subsection{Situation and Considerations}
Our design concept targets situations where pedestrians need to cross streets with heavy traffic and there are no pedestrian crossing facilities nearby. In such situations, where pedestrians tend to jaywalk, real-time safety information can be crucial to avoid accidents and to enhance the perceived safety and experience for pedestrians.

For the development of our design concept, we considered four types of safety-related information: permissive alerts, prohibitive warnings, directional warnings, and collision emergency warnings. Permissive alerts indicate when it is safe to cross, guiding pedestrians to choose optimal timing and/or paths for crossing, thereby reducing uncertainty and enhancing safety~\cite{malik2021determining}. Prohibitive warnings signal when it is unsafe to cross, preventing pedestrians from stepping into hazardous situations~\cite{malik2021determining}. Directional warnings notify pedestrians of approaching vehicles from the left or right, allowing them to better assess traffic flow and avoid collisions~\cite{hollander2020save}. Lastly, collision emergency warnings provide high-priority signals (for example through visual and audio cues) to urge pedestrians to stop or accelerate their crossing to avoid imminent danger~\cite{hollander2020save}.

The rationale behind focusing on these warnings stemmed from the need to address the core challenges of pedestrian safety in uncontrolled street-crossing environments. The ability to communicate this information clearly and in real-time was a primary consideration throughout the design process.

\begin{figure}[htbp]
\centerline{\includegraphics[width=\linewidth]{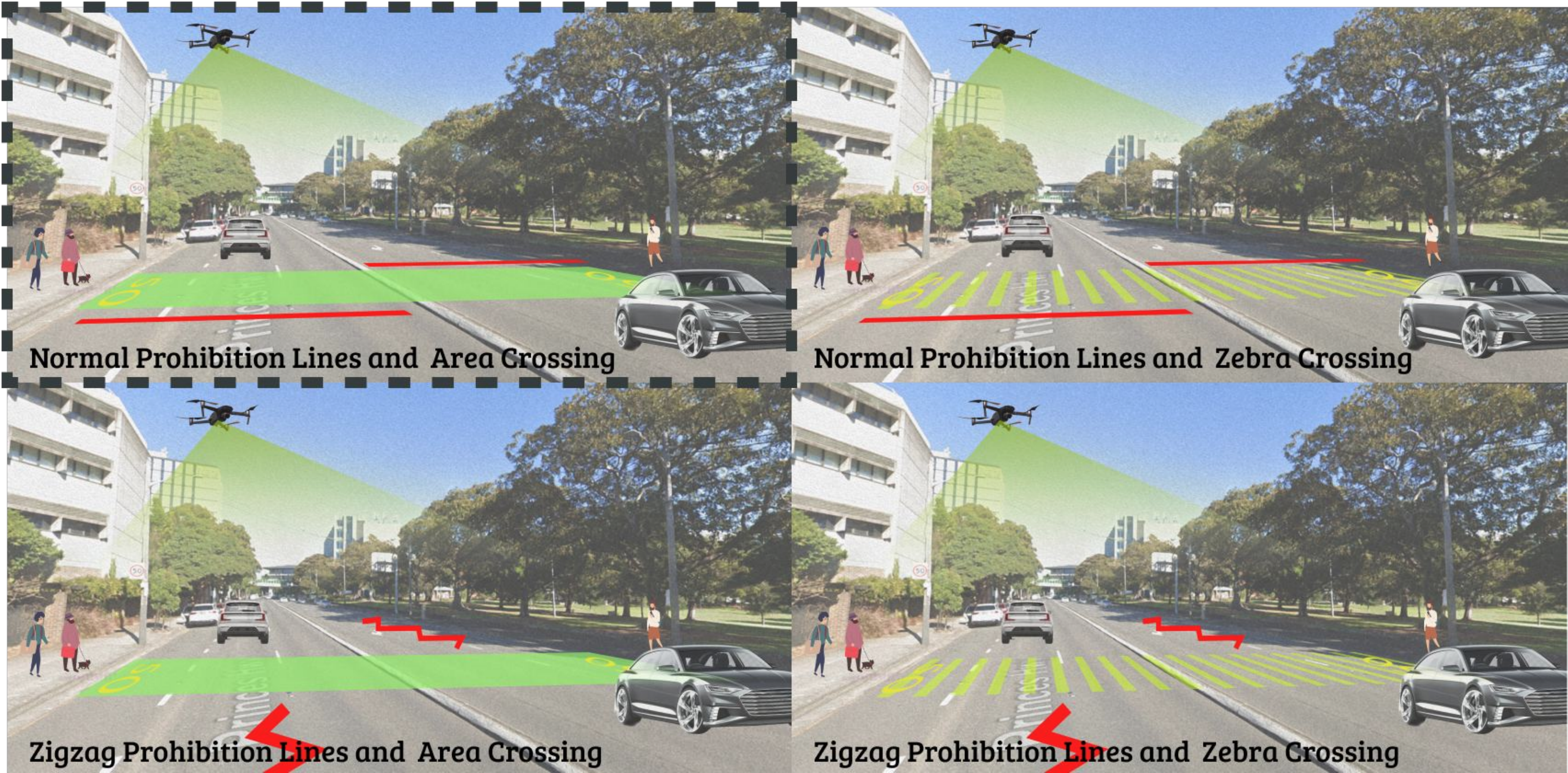}}
\caption{Initial design ideas for safety-cue visualisation.}
\label{ideas}
\end{figure}

\subsection{Initial Design Concept}
Our initial design concept focused on the use of in-situ projections as the primary method for communicating safety information to pedestrians. ``In-situ'' denotes the use of projections to display information directly in the physical environment~\cite{knierim2018quadcopter}. Making use of a drone's mobility to follow or stay in proximity of pedestrians, along with its ability to dynamically track real-time traffic data~(e.g.,~through computer vision or in combination with other V2X infrastructure), the safety-related information can be directly embedded on top of their corresponding physical point of reference~\cite{Suzuki2022, Willett2017}. We prototyped initial design ideas for visualisation and interaction in 2D, using footage from Google Maps for background imagery and incorporating a DJI Mavic 2 drone image~(see~Figure~\ref{ideas}).

\begin{figure}[hbt!]
\centerline{\includegraphics[width=0.5\linewidth]{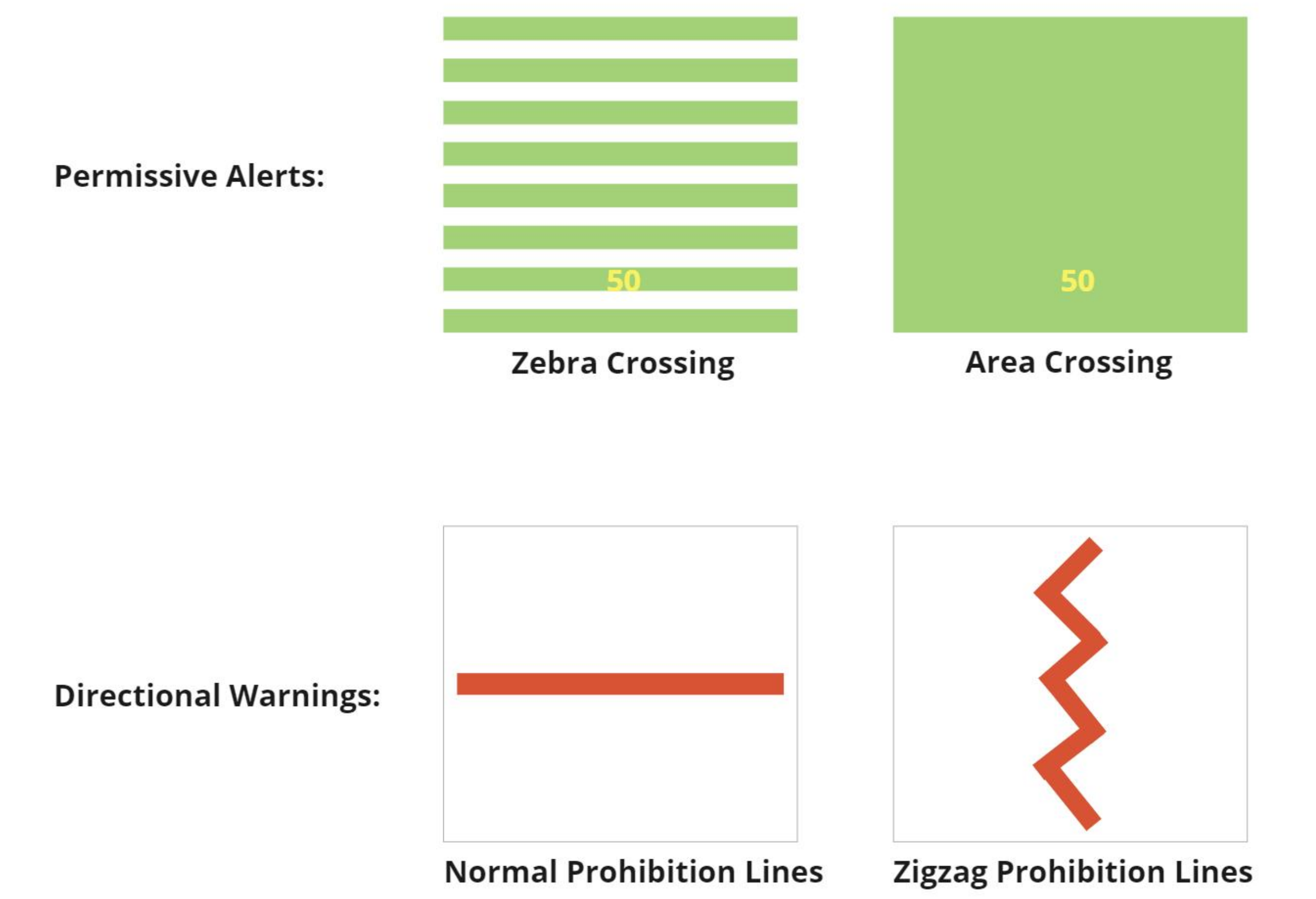}}
\caption{Permissive alerts and directional warnings.}
\label{alerts}
\end{figure}

Our initial design drew inspirations from symbols and colours found in conventional pedestrian crossing facilities, including crosswalks, pedestrian traffic lights, and crossing countdown timers(see~Figure~\ref{alerts}). Green and red were used to symbolise permission and prohibition for pedestrians to cross. A designated crosswalk area indicated a safe zone for crossing. While prohibition lines are traditionally used for guiding vehicles (e.g., slowing down)~\cite{NSW2021}, we adapted this design to assist pedestrians in effectively directing their attention to the right or left sides of the road, thereby mitigating the potential hazard of colliding with oncoming traffic.

\begin{figure*}[htbp]
\centerline{\includegraphics[width=\textwidth]{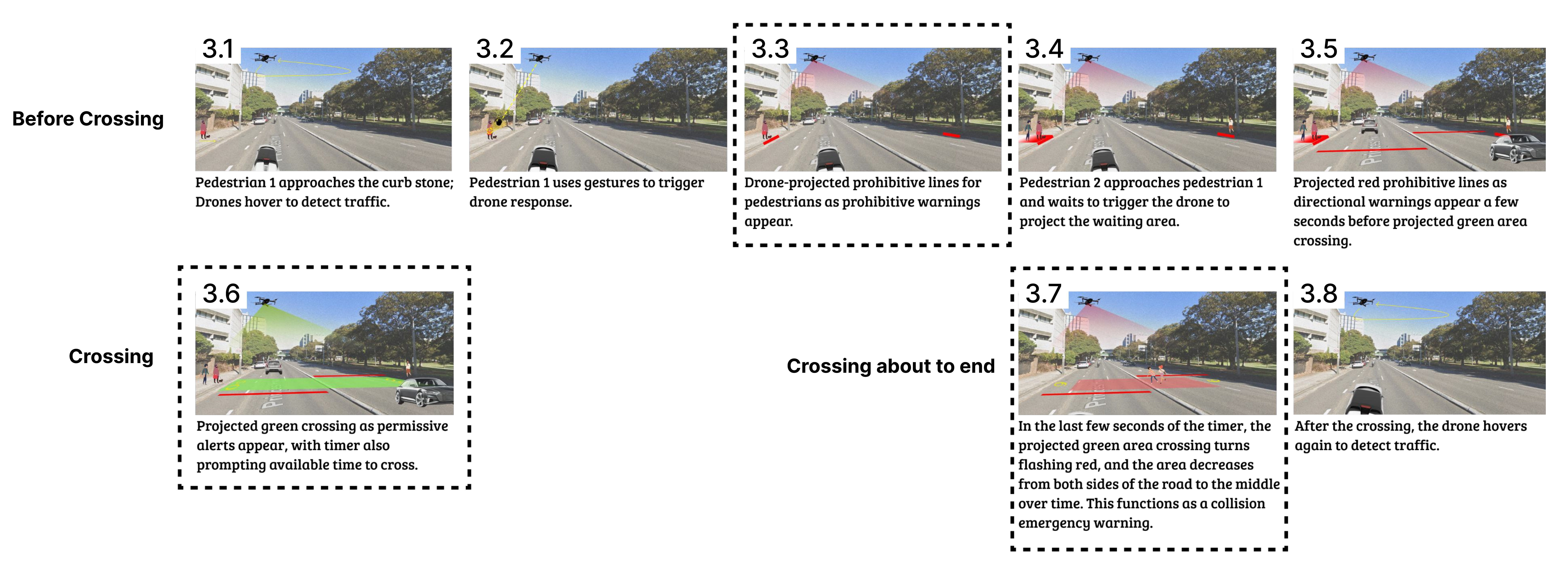}}
\caption{Storyboarding for ``before-crossing'', ``crossing'', and ``crossing-about-to-end''.}
\label{storyboard}
\end{figure*}

\begin{figure}[htbp]
\centerline{\includegraphics[width=0.7\linewidth]{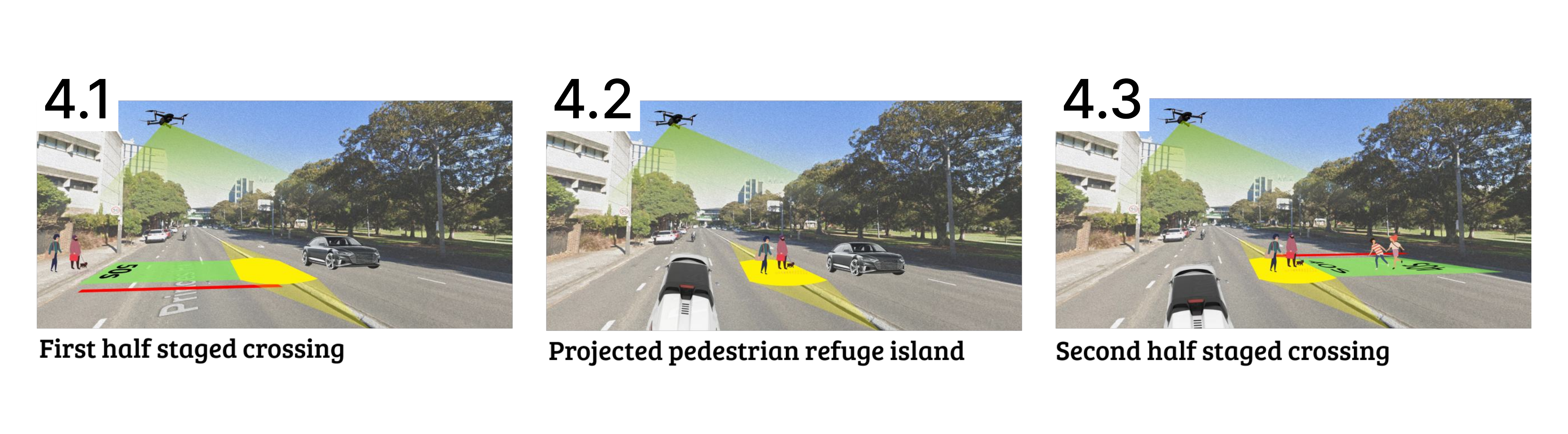}}
\caption{Staged Crossing, a part of the initial design for ``crossing''.}
\label{staged}
\end{figure}

\begin{figure*}[htbp]
\centerline{\includegraphics[width=\textwidth]{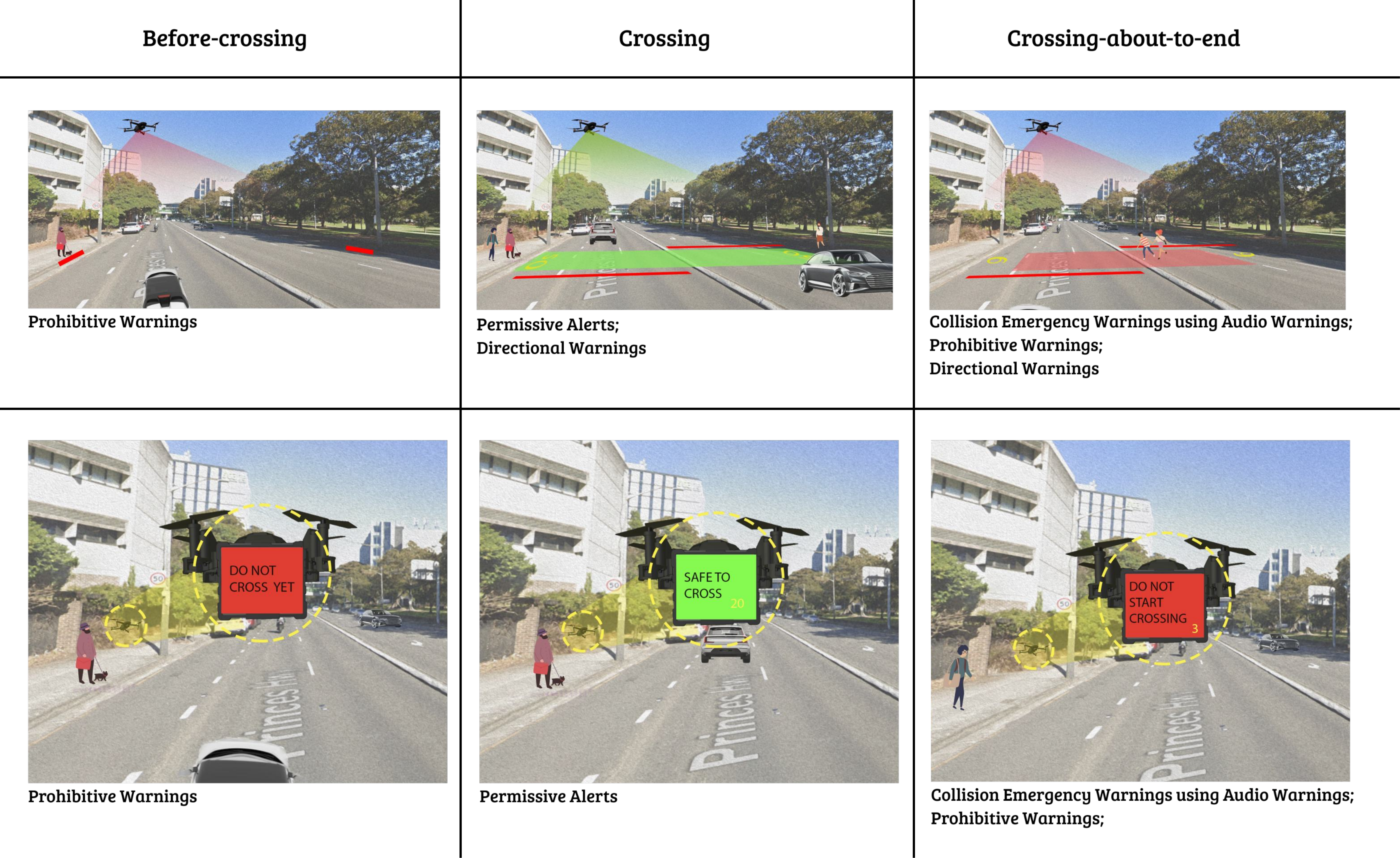}}
\caption{Drone with in-situ projections (top) and drone-equipped screens (bottom) that convey safety information for pedestrian crossing.}
\label{designs}
\end{figure*}
\raggedbottom

Building on these design elements, we initially came up with four design ideas resulting from a combination of two crosswalk styles (zebra crossing, area crossing) and two prohibition line styles (normal line, zigzag line), presented in Figure~\ref{ideas}. Additionally, we added a yellow countdown timer to these designs to provide pedestrians with information about the remaining safe crossing time~(making use of time-to-collision estimations based on the drone's real-time data collection).

Subsequently, we created a storyboard to map out the interactions of a pedestrian (intending to cross) with the drone-assisted system. Figure~\ref{storyboard} presents the interaction in three phases: ``before-crossing'', ``crossing'', and ``crossing-about-to-end''. The storyboard helped us to further think through key moments in the interaction: for example, we considered the use of gestures for the pedestrian to trigger initial engagement with the drone or a flashing red area to further emphasise the urgency in the final moments of crossing. In addition, we extended key frame 3.6 to explore a ``staged crossing'' concept tailored for multi-lane roads with a refugee island as a safe midpoint~(Figure~\ref{staged}).

\subsection{Design Refinement}
To understand the feasibility and clarity of the initial design and allow for further refinement, we conducted a feedback session with eight domain experts in robotics and interaction design, including five senior researchers and three PhD students specialising in HRI and AV-pedestrian interaction. The open discussion, augmented by the 2D prototypes, revealed several key insights. First, experts recommended simplifying the design to focus on its core functionality, such as conveying safety information, rather than including stylistic variations. Second, the staged crossing concept, while acknowledged, was considered too complex for the early proof-of-concept and hence was excluded from further iterations. Finally, experts suggested incorporating a screen-based interface as an alternative to in-situ projections, due to its popularity in HDI and being more practical than the projections.

Based on the above feedback, two final drone-assisted systems for pedestrian crossing were developed (Figure~\ref{designs}). The first system employed in-situ projection, providing visual cues directly onto the road surface. A green crossing area represented the safe zone for pedestrians, with a countdown timer showing the remaining safe crossing time. Red prohibition lines served as directional warnings, guiding both pedestrians and vehicles. In the final seconds of the crossing, the projected area flashed red to indicate urgency, along with a ``beep'' audio warning (synchronised with the visual flashing interval). 

The second system used a drone-equipped screen to display the textual messages ``safe to cross'', ``do not cross yet'', and ``do not start crossing'' with a green or red background to indicate the safety status. The use of textual messages is a common approach in AV-pedestrian interaction~\cite{bazilinskyy2019survey}, and text-based eHMIs have been shown to be highly comprehensible compared to other modalities~\cite{chang2018video}. Same as for the in-situ projections, we also incorporated a countdown timer to indicate the time remaining to cross as well as visual cue and audio warning~(screen flashing red and ``beep'' sound) to emphasise final seconds of safe crossing.

While these two systems represent different interface modalities, the safety cues were designed based on a similar rationale for addressing the same safety challenges. This allows for a direct comparison between the two systems, providing deeper insights into the effectiveness of drone interfaces for pedestrian crossings and which interface modality may be preferred in various crossing scenarios.

\begin{figure}[htbp]
\centerline{\includegraphics[width=0.65\linewidth]{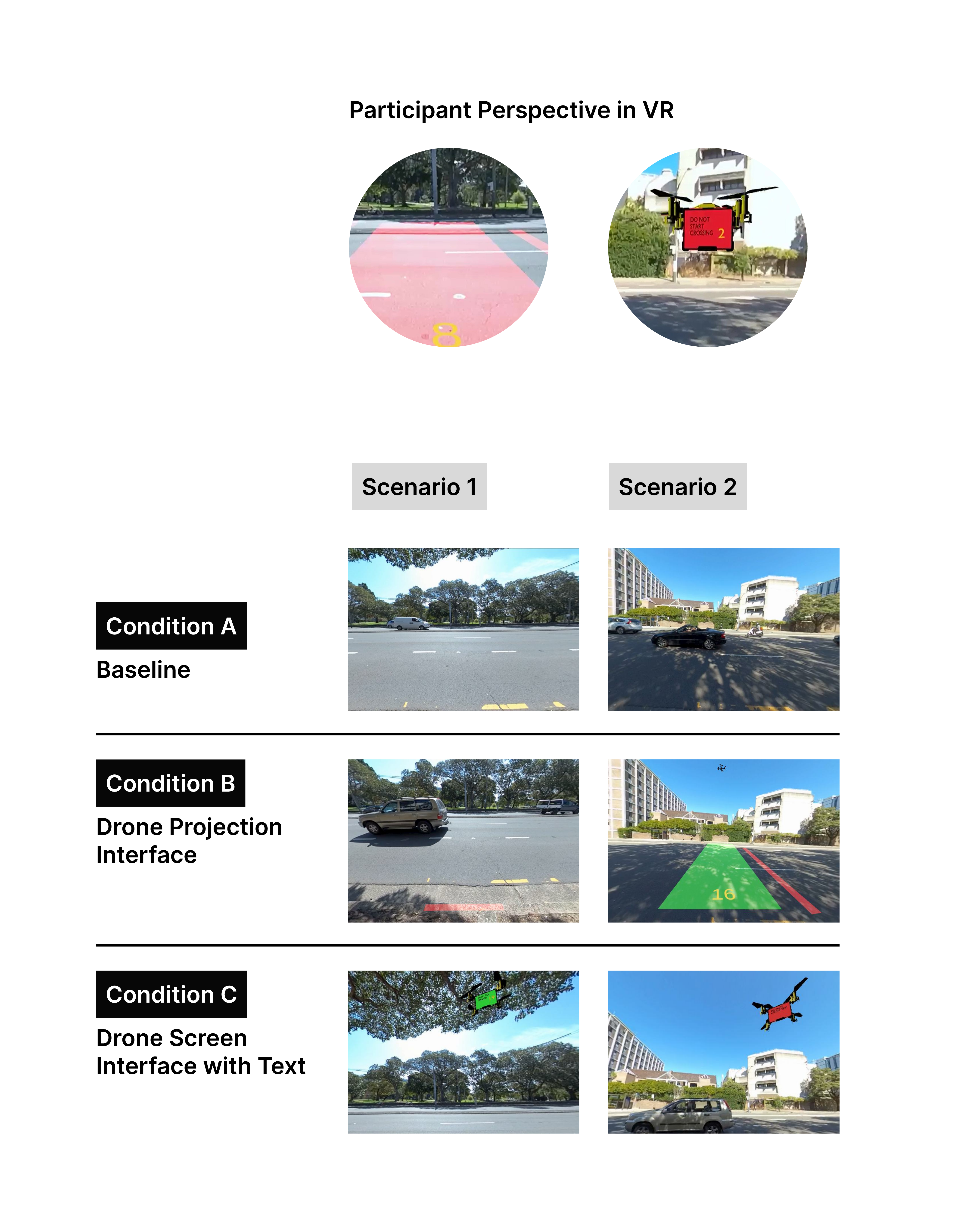}}
\caption{Experimental design for 3 (conditions) x 2 (types of scenarios).}
\label{evaluation}
\end{figure}

\section{User Evaluation}

\subsection{Experiment Design}
To facilitate comparative analysis, a baseline condition, which included no drone assistance, was introduced alongside the two drone-assisted systems~(see Figure~\ref{evaluation}). This design allowed for a direct comparison of pedestrian behaviour and perceptions under three distinct conditions: drone with in-situ projections, drone-equipped screens, and the baseline without any assistance.

\begin{figure}[htbp]
\centerline{\includegraphics[width=0.7\linewidth]{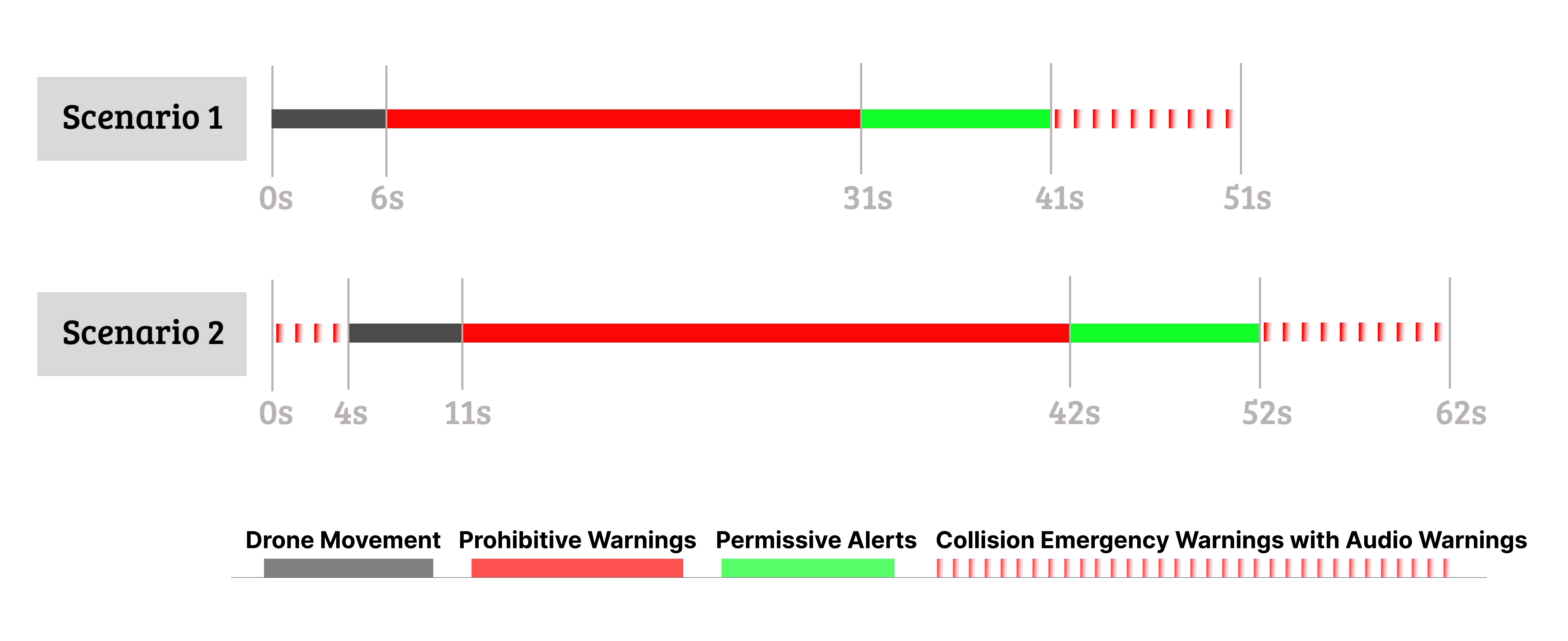}}
\caption{Configuration of animation for two distinct scenarios.}
\label{scenario}
\end{figure}

The evaluation incorporated two separate road-crossing scenarios. These scenarios were carefully selected from our 360-degree recorded footage to simulate real-world conditions in which pedestrians must cross busy roads without designated crossing facilities. The footage was recorded on a city road characterised by its wide, two-way, four-lane layout. Situated in a bustling urban neighbourhood, the road and its neighbouring pavements are frequented by various vulnerable road users. The street presents significant challenges for safe crossing: pedestrian facilities are sparse, with the closest designated crossing areas approximately 190 meters apart. Key facilities, such as bus stations and several university buildings situated between these crossing, further increase pedestrian activity, making the area highly prone to jaywalking. Additionally, trees along one side of the road obstruct visibility, making it difficult for pedestrians to see both the crossing facilities and oncoming traffic from the recording location.

Two scenarios each featured a forth or back crossing this road between point A and point B. So, the pedestrian stood on point A and faced point B (Scenario 1), and vice versa (Scenario 2). In Scenario 1, participants observed traffic conditions and decided when it was safe to cross the street. In Scenario 2, participants began with the tail end of a previous crossing gap, making judgement about whether it was safe to continue crossing in the heavy traffic. Figure~\ref{scenario} shows the configuration of the two scenarios. Each scenario was presented under the three experimental conditions, resulting in six distinct situations that participants encountered (Figure~\ref{evaluation}). We 3D modelled the two drone-assisted systems in Blender. We then imported both the 3D models and the video scenarios into Unity, and developed a VR simulation combining both materials. The simulation was deployed to Oculus Quest 2. Users indicated their decision to cross by pressing the trigger button on the right controller.

\subsection{Procedure}
Upon arrival, participants were briefed on the study's  background and procedures. They were provided with two documents: a participant information statement, which included detailed information about the research, and a consent form. The consent form gave participants the opportunity to agree to audio and video recording and the use of their photographs. After signing the consent form, participants completed a demographic questionnaire covering age, gender, occupation, and prior experience with VR and AVs.

The order of conditions was randomised using a Balanced Latin Square design to minimise carryover effects in this within-subjects lab study. Participants used the VR headset to engage with the VR simulation and indicated their decision to cross by clicking the trigger button on the right controller. After completing each condition, they were asked to fill out the questionnaires to assess their perceived safety, trust, mental workload, and user experience (UX) of the systems. Following the completion of all conditions, participants undertook a short semi-structured interview to provide qualitative insights.

\subsection{Participants}
The user study was conducted with eighteen participants~(13 men, 5 women) between the ages of 20–34 years (M=24.8, SD=3.0), recruited through posters and mailing lists. Most participants reported infrequently using VR, three had no prior VR experience, and two reported frequently using VR. Familiarity with AVs was similarly limited, with only two participants reporting frequent interaction with AVs. All participants had normal or corrected-to-normal vision to ensure that they could fully engage with the VR simulation. The study adhered to the ethical approval from the human research ethics committee of University of Sydney.

\subsection{Data Collection}
To gain a comprehensive understanding of the systems' impact, the study evaluated perceived safety~\cite{bartneck2009measurement}, trust~\cite{jian2000foundations}, mental workload (NASA-TLX), UX of the systems (AttrakDiff)~\cite{de2021intelligent} using validated questionnaires, as well as the overall crossing experience using semi-structured interviews.

Perceived safety was assessed with semantic differential scales ranging from -3 to 3, which is tailored for evaluating the perceived safety of robotics~\cite{bartneck2009measurement}. The Trust in Automated Systems scale measured both trust and distrust on a 7-point Likert scale~\cite{jian2000foundations}. Mental workload was assessed with a benchmark scale ranging from 1 to 20 (NASA-TLX). UX was measured via four dimensions: pragmatic quality (PQ), hedonic quality – identification (HQI), hedonic quality – stimulation (HQS), and attractiveness (ATT)~\cite{de2021intelligent}. Each dimension was measured on a 7-point scale ranging from -3 to 3, designed to assess UX of interactive systems.

\subsection{Data Analysis}
Quantitative data analysis began with a descriptive analysis of the questionnaire responses, followed by non-parametric statistical tests. The Friedman test was used to assess the statistical significance of differences across the three conditions. Where the Friedman test indicated significant differences, Wilcoxon signed-rank tests were conducted to identify specific pairwise differences between conditions. A p-value of less than 0.05 was considered statistically significant.

Qualitative data analysis involved transcribing the interview recordings and applying affinity diagramming to categorise and cluster similar concepts. These concepts were examined in order to contextualise and deepen the interpretation of quantitative findings.

\begin{figure*}[htbp]
\centerline{\includegraphics[width=\textwidth]{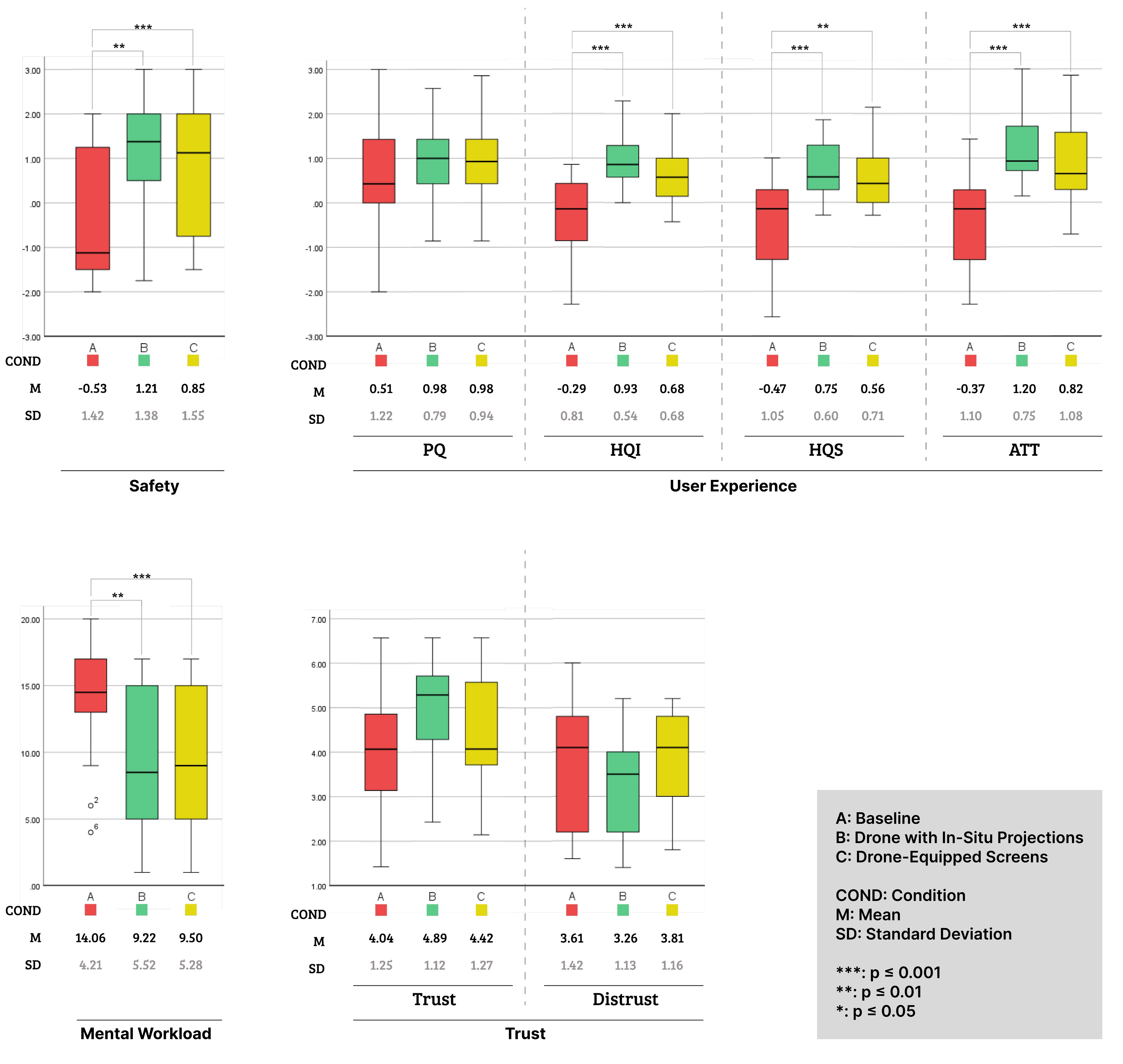}}
\caption{Box plots of test results for perceived safety, trust, mental workload, and user experience, with means (M) and standard deviations (SD). Significant differences between conditions are indicated with asterisks: * p $<$ 0.05, ** p $<$ 0.01, *** p $<$ 0.001. }
\label{stats}
\end{figure*}

\section{Results}

\subsection{Perceived Safety}

\subsubsection{Perceived Safety Scale}
Descriptive analysis (see Figure~\ref{stats}) indicates that among the three conditions, Condition B (drone with in-situ projections) yielded the highest perceived safety rating. Condition C (drone-equipped screens) followed closely, though with a slightly lower rating than Condition B. In contrast, Condition A (baseline without crossing support) showed a substantially lower safety rating compared to Condition B. Results from the Friedman Test revealed a statistically significant difference in perceived safety across the conditions, {$\chi^2$} = 13.268, p = 0.001. Post hoc analysis using Wilcoxon signed-rank tests showed that both Condition B (p = 0.003) and Condition C (p $<$ 0.001) had significantly higher safety ratings than Condition A.

\subsubsection{Qualitative Feedback on Perceived Safety}
During the interviews seven participants stated that the drone-assisted systems made them feel safer compared to the baseline condition, aligning with the statistical findings. Five participants attributed this to the drones' ability to simulate traffic signals, such as time countdowns. One participant referred to the potential of drones to ubiquitously act as ``movable traffic lights'' that provide more ``safety signals or safety concerns'' for crossing. Additionally, two participants appreciated how drones communicated intuitively through their movement, forming a temporary relationship with them during the task of safely crossing the road.

\subsection{Mental Workload}

\subsubsection{Mental Workload Scale}
In terms of mental workload (see Figure~\ref{stats}), Condition A (baseline) exhibited the highest rating, while Condition C (drone-equipped screen) showed a much lower mental workload. Condition B (in-situ projection) also demonstrated a lower rating compared to Condition A, though slightly higher than Condition C. The Friedman Test revealed a significant difference in mental workload across the three conditions, {$\chi^2$} = 11.030, p = 0.004. Post hoc analysis using Wilcoxon signed-rank tests showed that both Condition B (p = 0.003) and Condition C (p $<$ 0.001) had significantly lower mental workload ratings compared to Condition A.

\subsubsection{Qualitative Feedback on Mental Workload}
Five participants felt using drone-assisted systems reduced their mental workload compared to the baseline. This aligns with the quantitative results that showed a significant reduction in mental workload in the drone-assisted conditions. Three participants mentioned that the drones simplified the crossing process, with one saying they ``did not need to look left and right'' to gather safety information. Three participants expressed high reliance on the systems, while three others viewed it more as a ``double-check'' mechanism. Additionally, three participants felt the drone systems reduced the attention needed when crossing, with one noting it could help ``save time.''

\subsection{Trust}

\subsubsection{Trust Scale}
Descriptive analysis (see Figure~\ref{stats}) shows that Condition B (in-situ projection) had the highest trust rating and the lowest distrust rating. Trust and distrust levels for Condition A (baseline) and Condition C (drone-equipped screen) were similar, with Condition C showing a slightly higher trust rating. However, the Friedman Test results showed no statistically significant differences in trust ({$\chi^2$} = 5.768, p = 0.056) or distrust ({$\chi^2$} = 3.794, p = 0.150) across conditions.

\subsubsection{Qualitative Feedback on Trust}
Despite the lack of significant statistical differences, eleven participants expressed higher trust in the drone-assisted systems compared to the baseline. Nine participants identified the ``traffic signal simulation'' as the main reason for their enhanced trust, with six noting that drones made traffic safety data more predictable. Three participants appreciated how the in-situ projections displayed safety information in a ``straightforward manner'' by directly overlaying it onto the physical space relevant to the crossing task. Additionally, six participants felt that the movement of the drones increased their sense of trust, with three describing the drones as ``detectors'' when hovering around to gather safety-related information. However, five participants expressed distrust, with three specifically concerned about the risk of bodily injury due to the close proximity of the drones. One participant further raised concerns about drones colliding with vehicles, and one participant remarked that the projections felt too novel and therefore unfamiliar, making it hard to build trust within a short period.

\subsection{User Experience}

\subsubsection{AttrakDiff Questionnaire}
Descriptive analysis (see Figure~\ref{stats}) shows that Condition B (in-situ projection) scored the highest for all UX factors, including Pragmatic Quality (PQ), Hedonic Quality--Identification (HQI), Hedonic Quality--Stimulation (HQS), and Attractiveness (ATT). Condition C (drone-equipped screen) followed closely behind, with Condition A (baseline) scoring significantly lower for all factors. Friedman Test revealed statistically significant differences in HQI ({$\chi^2$} = 19.906, p $<$ 0.001), HQS ({$\chi^2$} = 13.457, p = 0.001), and ATT ({$\chi^2$} = 19.127, p $<$ 0.001). Wilcoxon signed-rank tests showed both Condition B (p $<$ 0.001) and Condition C (p $<$ 0.001) had significantly higher HQI ratings than Condition A. Similarly, Condition B (p $<$ 0.001) and Condition C (p = 0.005) had significantly higher HQS ratings compared to Condition A, and Condition B (p $<$ 0.001) and Condition C (p = 0.001) showed significantly higher ATT ratings than Condition A.

\subsubsection{Qualitative Feedback on User Experience}
Five participants reported a better overall experience with the drone-assisted systems compared to the baseline, consistent with the quantitative findings. Three participants expressed interest in how the drones used their movement for safety notifications, with one even expressing to have felt a sense of connection with the drone. Two participants reported feeling more 
relaxed when using the drone systems. However, one participant criticised the drone model, describing it as too ``industrial''-looking, and suggested that a more ``cute and friendly'' design could enhance experience.

\section{Discussion}

\subsection{Drone Interfaces as Distributed Traffic Light Systems}
Pedestrian crossing at uncontrolled intersections or along roadways without nearby road crossing facilities is a critical safety issue due to the lack of traffic moderation~\cite{rasouli2019autonomous}. This scenario is a primary focus in the design of negotiation strategies for AVs, where eHMIs play a vital role in coordinating movement between pedestrians and vehicles in the absence of traditional traffic signals~\cite{dey2020taming}.

As cities move towards greater digitisation and connected transportation networks, eHMIs are evolving beyond vehicle-attached systems. Proposals increasingly suggest that they can also function as third-party mediators, acting as proxies for vehicles and other road users through V2X technology~\cite{kavas2022v2x}. Several infrastructure-based solutions, such as smart roadside poles~\cite{chauhan2023fostering}, smart kerbs~\cite{hollander2022take}, and smart roads~\cite{smartRoad2017}, which gather, interpret, and relay traffic information, are already being tested in both simulation-based studies and testbed implementations to communicate traffic data to both vehicles and pedestrians. Third-party eHMIs can also include personal devices like augmented reality glasses~\cite{tran2022designing} or smartphones~\cite{hollander2020save}, which allow pedestrians to communicate with AVs directly. Our drone proposals are conceivable because increasing technical and hardware studies point to the promising future of UAV transceivers communicating with automated vehicles via 5G networks~\cite{kavas2022v2x,cao2022toward}. A mobile projection-based approach supports flexibility and particularly laser projections are promising as they are both visible and colourful, with real-world application in dynamically alerting and guiding cyclists~\cite{dancu2015gesture}. This decentralised approach to eHMIs supports personalisation and individual crossing intentions.

In our study, which is the first to empirically test a mobile drone-assisted crossing system, participants likened the design concept to ``traffic light systems''. The similarity could be due to the interface designs being inspired by traditional crossing facilities; more importantly, in non-signalised road scenarios where pedestrians lack clear support, the drones acted as mediators, providing real-time traffic status updates. This increased situational awareness allowed pedestrians to make more informed crossing decisions. Participants viewed the drones as a ``double-check'' system, reinforcing their judgement and boosting trust by clearly conveying safety data, thus reducing mental workload and enhancing the overall experience. Meanwhile, it is important for designers and engineers to consider the potential risk of overtrust in such automated systems, particularly in the event of technical failures~\cite{khastgir2018calibrating} which can reduce pedestrians’ situational awareness of traffic dangers. To mitigate this, we recommend implementing measures to calibrate pedestrian trust in automated agents~\cite{m2021calibrating}. For example, downplaying the agent's `autonomous' status and instead emphasising intent messages (e.g. ``go'' or ``not go'') has proven effective in reducing overtrust and improving crossing safety with automated vehicles~\cite{m2021calibrating}.

Our findings suggest the potential for drone interfaces to function as distributed traffic light systems. The mobility and expressiveness of drones can be harnessed to communicate traffic information effectively, while also addressing individual road-crossing needs. Looking forward, drones could evolve into public assets integrated into V2X infrastructure~\cite{herdel2021public,kavas2022v2x}, offering scalable, adaptable support for pedestrians in diverse traffic environments.

\subsection{Perception of Connectedness Varied in Drone Interfaces}
The drone interfaces played an intermediary role in the interaction between pedestrians and vehicles within a connected, dynamic traffic environment. The drone-equipped screens moved along with the drone and were well-received by participants for fostering a sense of engagement and trust, creating connectedness between the pedestrian and the drone by operating within personal distance of proxemics~\cite{hall1968proxemics}. Participants felt more engaged with this system as the drone was in closer range, with its dynamic movement capturing their attention and heightening their sense of safety. However, maintaining an appropriate distance was noted as crucial, as excessive proximity could raise concerns about bodily harm.

Conversely, participants who preferred the projection system appreciated the drone’s greater distance, which helped alleviate concerns about physical safety. The drone’s hovering presence, perceived as effectively monitoring traffic, reinforced trust in its ability to enhance safety. Since participants valued vehicles being aware of their crossing behaviour, the larger, more visible projection fostered a sense of shared awareness between pedestrians, vehicles, and the drone, as drivers or oncoming AVs could also see the same signals projected onto the road. This increased connectedness among all traffic participants further strengthening the perception of safety.

These findings highlight the importance of designing drone interfaces that balance proximity and visibility to create meaningful connectedness that support both technology-user relationships and shared awareness among broader road users. This adds a new perspective to the role of proxemics in human-robot interaction beyond single-person scenarios~\cite{electronics11162490, LEICHTMANN2020101386}.

\subsection{Design Recommendations for Drone Interfaces to Assist Safe Pedestrian Crossing}

Based on our findings, we outline the following design recommendations (DR.) for designing drone interfaces that support road-crossing.

DR.1--Mimic traffic signals: To improve public understanding of drone-assisted crossing systems, incorporating familiar traffic signals can support intuitive and safe experiences.

DR.2--Consider using in-situ projections:
Projecting safety information onto road surface enhances visibility and intuitiveness for pedestrians. It also helps vehicles detect pedestrians more easily, contributing to a stronger sense of safety.

DR.3--Consider proximity of the drone:
An interpersonal proximity between the drone and the pedestrian can establish connection, trust, and a sense of companionship, while still requiring a balance with safe operational distance.

DR.4--Combine visual and audio designs:
Auditory cues can effectively complement visual signals, particularly for permissive and prohibitive alerts, providing another layer of safety message communication for pedestrians.

DR.5--Utilise drone movements for status report:
Drone movements can signal its operational status, such as hovering to indicate traffic monitoring or approaching to signal readiness for crossing. This dynamic interaction strengthens trust and captures attention.

\subsection{Limitations and Future Work}
We used real-world 360-degree camera footage to simulate a busy roadway, increasing the realism of danger and providing a more realistic context~\cite{hoggenmuller2021context,wang2024immersive} for testing the drone interfaces. However, the real-world sound in the footage was not spatialised. Spatial sound is found to significantly enhance the presence and realism in VR~\cite{Mavridou2025SpatialAudioVR,10.3389/fpsyg.2021.628298}. Future VR evaluations of drone-crossing interfaces could integrate spatialised audio (e.g., directional engine) to more fully simulate real-world experiences.

While VR does not fully capture all nuances of real-world interactions, it provides a controlled, immersive environment that captures specific user interactions of interest in risky scenarios without endangering participants. Studies have shown that VR provides acceptable ecological validity for AV-pedestrian~\cite{tran2024advancing,song2024road} and HRI research~\cite{yu2023your,mara2021user}, making it a practical and effective first step toward real-world application. Regulatory and safety constraints make testing in live traffic environments challenging, and even dedicated road safety test bed environments, while valuable, are costly and cannot replicate the complex nature of urban situations. Using 360-degree camera recordings in VR instead enabled us to simulate realistic urban scenes cost-effectively. Given the pressing need for pedestrian safety solutions, we hope that our proof-of-concept in VR will inspire further research and facilitate the transition to real-world testing to tackle this urgent safety issue.

We employed a within-subjects design to control for individual differences and enhance internal validity and to enable participants to provide comparative feedback across interfaces, enriching our qualitative data. To manage potential order effects, we counter-balanced interface presentations in our study; however, a between-subjects approach could further reduce order effects, albeit requiring a larger sample size for comparable statistical power and risking noise via individual variability~\cite{hoffman2020primer}. Most in-person HRI studies use sample sizes below 25 participants~\cite{zimmerman2022analysis}, which is also a common and practical approach for mixed-methods VR lab studies~\cite{yu2023your,colley2022effects,wang2025passersby}. As an exploratory study in drone-assisted pedestrian safety, our work provides a foundation for future work, which could expand sample sizes, explore between-subjects designs, diversify participant profiles, and consider varied traffic scenarios and cultural contexts to advance the generalisability of our designs.

Other forms of infrastructure-based crossing supports, such as a smart-pole interaction unit~\cite{inproceedings}, were not compared to our drone-based solutions because we focused on the drone interfaces to clearly see their impact. Future studies should conduct experiments to further understand effects of drone-assisted crossing facilities compared with other V2X infrastructure.

\section{Conclusion}
We designed drone interfaces--using in-situ projections and drone-equipped screens--to assist pedestrians in crossing non-signalised roads. Through an iterative design process, we represented crossing-related safety cues by adapting design elements from existing traffic systems. In a VR evaluation with eighteen participants, projections outperformed screens, and both significantly improved perceived safety, mental workload, and the hedonic aspects of UX compared to no crossing aids. Our findings demonstrate the potential for future drone interfaces to act as distributed traffic light systems, mediating traffic by fostering connectedness among road users. With the pressing pedestrian safety issue on urban roads, we emphasise the urgency of proposing new solutions for smart traffic networks. Before V2X and AVs fully arrive, it is essential to evaluate possible designs that align with these technologies. Our work thus pioneers the use of drones for pedestrian safety and can spur future studies interested in this area to build on the insights gained. This paper particularly lays a design foundation for developing drone-assisted systems as scalable and distributed eHMIs to support safe pedestrian crossing on non-signalised roads.

\bibliographystyle{ACM-Reference-Format}
\bibliography{references}

\end{document}